\documentclass[journal,10pt]{IEEEtran}
\pdfoutput=1
\IEEEoverridecommandlockouts
\usepackage{amsbsy}
\usepackage{amsthm}
\usepackage{amssymb}
\usepackage{amsmath}
\usepackage{multirow}
\usepackage{epsfig}
\usepackage{subfigure}
\usepackage{graphics}
\usepackage{multicol}
\usepackage{diagbox}
\usepackage{upgreek}
\usepackage{threeparttable}

\usepackage{epstopdf}
\usepackage{balance}
\usepackage{xcolor}
\usepackage{cite}
\usepackage{color}
\usepackage{algorithm}
\usepackage{algorithmic}
\usepackage{textcomp}
\usepackage{subfigure}
\usepackage{subfloat}
\usepackage{booktabs}
\usepackage{bm}
\usepackage{filecontents}

\graphicspath{ {figures/}}

\begin{document}

\title{{\huge Sensing as a Service in 6G Perceptive Mobile Networks: Architecture, Advances, and the Road Ahead}}

\author{Fuwang Dong, Fan Liu, Yuanhao Cui, Shihang Lu, Yunxin Li

\thanks{The authors were with the Department of Electronic and Electrical Engineering, are now with the School of System Design and Intelligent Manufacturing, Southern University of Science and Technology, Shenzhen 518055, China. (\textit{Corresponding author: Fan Liu}, email: liuf6@sustech.edu.cn)}
}

\maketitle

\begin{abstract}
Sensing-as-a-service is anticipated to be the core feature of 6G perceptive mobile networks (PMN), where high-precision real-time sensing will become an inherent capability rather than being an auxiliary function as before. With the proliferation of wireless connected devices, resource allocation (RA) in terms of the users' specific quality-of-service (QoS) requirements plays a pivotal role in enhancing interference management ability and resource utilization efficiency. In this article, we comprehensively introduce the concept of sensing service in PMN, including the types of tasks, the distinctions/advantages compared to conventional networks, and the definitions of sensing QoS. Subsequently, we provide a unified RA framework in sensing-centric PMN and elaborate on the unique challenges. Furthermore, we present a typical case study named ``communication-assisted sensing'' and evaluate the performance trade-off between sensing and communication procedures. Finally, we shed light on several open problems and opportunities deserving further investigation in the future.     
\end{abstract}

\IEEEpeerreviewmaketitle

\section{Introduction}\label{Introduction}

\subsection{Background and Motivations: ISAC-Empowered 6G Perceptive Mobile Networks}

The recommendation proposal for IMT-2030 (6G) has been officially approved by the International Telecommunication Union (ITU) during the 44th meeting in Geneva in 2023. This marks a significant milestone in the development of the next-generation 6G wireless communication system. Unlike its predecessors, which mainly focused on achieving higher data rates and network capacity, the vision of 6G represents a radical paradigm shift to meet the requirements of emerging environment-aware applications, such as the Internet of Things (IoT), extended reality (XR) services (encompassing augmented, mixed, and virtual reality), metaverse, autonomous driving, and more. Driven by such applications, integrated sensing and communications (ISAC) along with native AI have become inevitable development trends to provide services beyond communications \cite{R1}.    

The ISAC methodology facilitates the sensing and communications (S\&C) functionalities by the effective shared use of spectral and power resources, hardware platform, and signal processing framework \cite{R2,RLiu}. Furthermore, various S\&C cooperation schemes have also been explored, enhancing the system's capabilities. For instance, in communication-centric scenarios, the sensing capabilities may assist in reducing the pilot training overhead for high mobility channel estimation \cite{R12}. Conversely, in sensing-centric scenarios, the communication function may be employed to sense the beyond-line-of-sight (BLoS) targets of interest, where the BLoS includes both ultra-long range and LoS-obstructed conditions \cite{R17}. The ISAC system thus offers significant improvements in S\&C performance, presenting new challenges and opportunities for system optimization compared to the current communication-only cellular network.  
  
Empowered by ISAC, current cellular network architecture is undergoing a paradigm shift, evolving into ubiquitously deployed large-scale sensor networks known as perceptive mobile networks (PMN) \cite{R3}. Both S\&C functionalities can be seamlessly implemented in a single node or across networked nodes, with only minor modifications in hardware, signaling strategy, and communication standards. These nodes can be either base station (BS) or user equipment. In the context of PMN, sensing-as-a-service emerges as the central feature, where accurate sensing is an inherent capability rather than an auxiliary function as before. In numerous intelligent applications, sensing quality-of-service (QoS) becomes equally, if not more, important compared to that of communications. This necessitates careful consideration of both S\&C QoSs during system design and resource scheduling.

\subsection{Physical Layer Resource Allocation for PMN Network}

 \begin{figure*}[!t]
	\centering
	\includegraphics[width=6.8in]{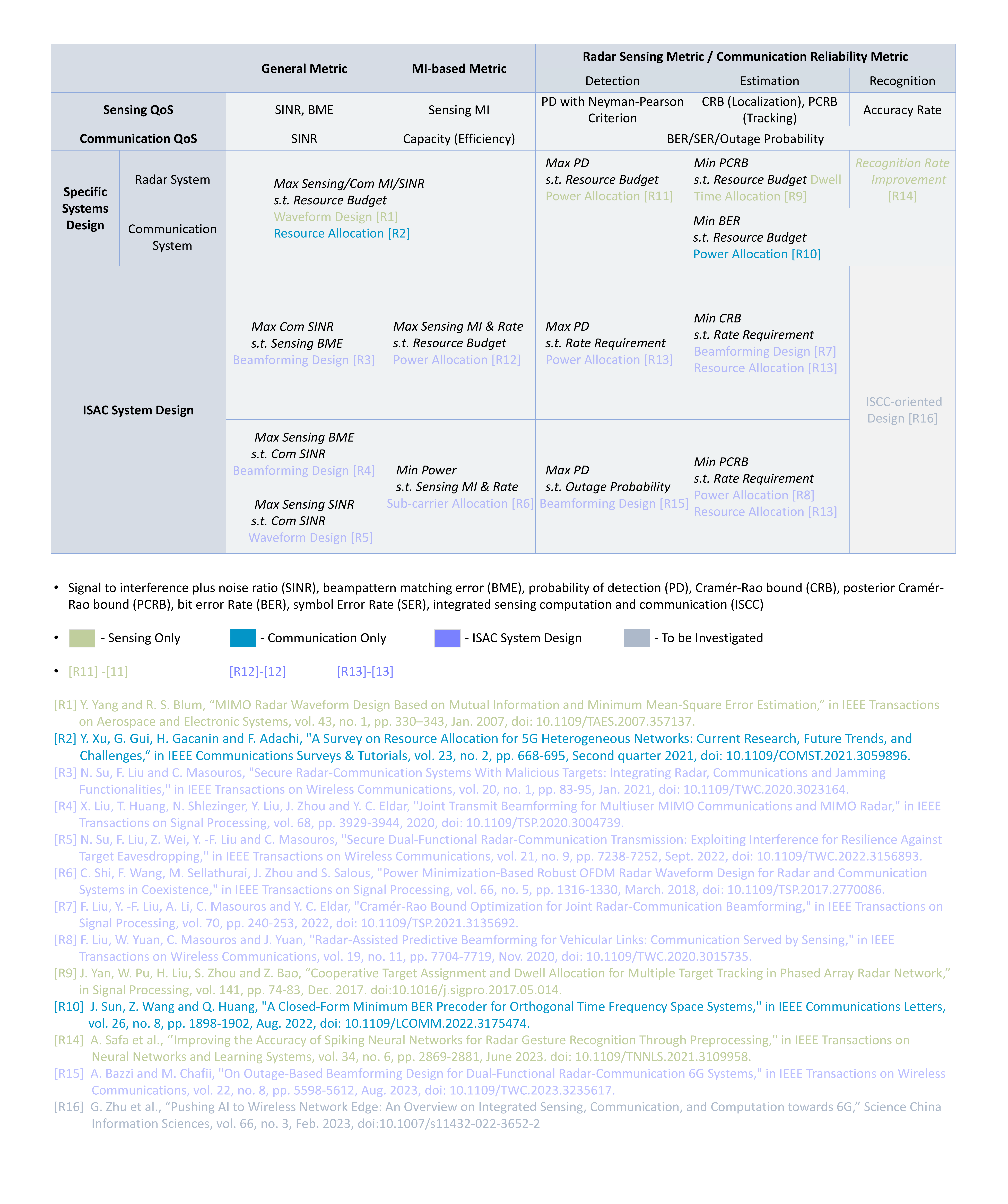}
	\caption{The S\&C performance metrics and the associated representative system design schemes.}
	\label{IR111}
\end{figure*}

Triggered by the development of numerous emerging applications, the proliferation of wireless connected devices exhibits an exponential trend. In such a case, resource allocation (RA) in terms of the specific QoS requirements of the users plays a pivotal role in enhancing the interference management ability and resource utilization efficiency. In what follows, we provide a brief review of the RA schemes for communication, radar, and ISAC systems, respectively. The resources related to the S\&C performance and the state-of-the-art RA schemes are summarized in Fig. \ref{IR111}.     

\subsubsection{RA for Communication Systems} The primary research focus of RA in communication systems includes power control and bandwidth/subcarrier allocation, which determines the achievable rate based on Shannon's theorem. For instance, the optimal joint power and subcarriers allocation scheme for orthogonal frequency division multiplexing (OFDM) systems has been investigated in \cite{R5}, where the concept of proportional fairness constraints was introduced to strike a balance between capacity and fairness. These conventional techniques have reached a relatively mature stage. On top of that, with the explosive growth of wireless devices, user scheduling/paring and antenna selection/beamforming can be regarded as a part of RA, whose aim is mitigating the mutual interference in heterogeneous network and downlink non-orthogonal multiple access (NOMA) systems \cite{R7}.            

\subsubsection{RA for Radar Systems} The primary function of a radar system is detection and estimation. The key resources that impact its performance include power, bandwidth, beams, and dwell time. In general, radar systems were rarely deployed in resource-scarce scenarios, thus the RA problem is not as urgent as that in communication systems. In \cite{R8}, the authors considered that the formed beams are insufficient for multi-target tracking, and developed a simultaneous multi-beam and power allocation scheme to enhance the worst-case tracking performance. More recently, a joint power and bandwidth allocation scheme has been investigated to minimize the posterior Cr\'amer-Rao bound (PCRB) for multiple targets in the tracking strategy \cite{R9}.  
 
\subsubsection{RA for ISAC Systems} There have been a number of studies exploring the RA problems in ISAC systems. In \cite{R10}, the authors investigated a dual-functional radar and communication transmitter that supports both communication users and radar receivers. The probability of detection is maximized through appropriately allocating transmit power between radar waveform and information signal, while meeting the information rate requirement. By adopting the conditional mutual information (MI) between the random target impulse response and the received signal as the sensing performance metric, the authors in \cite{R11} have developed a power allocation scheme for an OFDM-based ISAC system. As a step further, the work of \cite{R14} proposed a joint power and bandwidth allocation scheme tailored for specific sensing tasks, including detection, localization, and tracking while ensuring a certain level of communication QoS. Moreover, in the context of ISAC-empowered connected and autonomous vehicle (CAV) networks, \cite{R12} introduced a real-time beamwidth adjustment mechanism to track the communication receiver installed on an extended target-modeled vehicle, facilitating the establishment and maintenance of communication links.   

In this article, we attempt to contribute to the concept of sensing-as-a-service in 6G PMN including its fundamental architecture, the unified sensing-centric RA framework, and the future research directions. We start by comprehensively introducing the concept of sensing service in PMN, encompassing the types of tasks, the differences/advantages compared to conventional networks, and the definition of sensing QoS. Subsequently, we provide a unified RA framework in sensing-centric PMN and elaborate on the unique challenges. Furthermore, we study a typical use case named ``communication-assisted sensing'' for the PMN. Finally, we shed light on several open problems and opportunities deserving further investigation in the future.

\section{Sensing as a Service in PMN}
  
\subsection{What is Sensing Service?}
\begin{figure}[!t]
	\centering
	\includegraphics[width=3.5in]{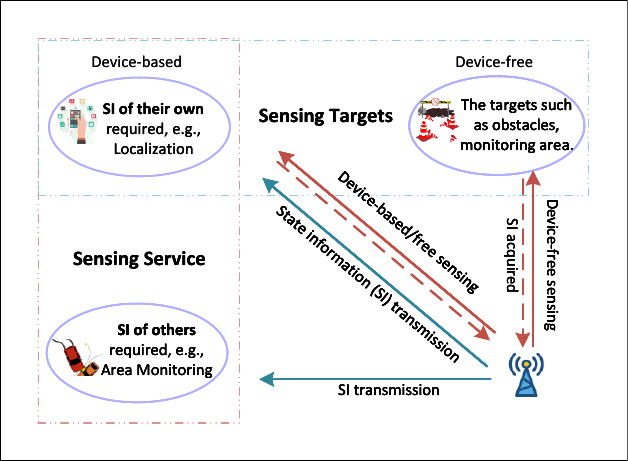}
	\caption{The types of sensing service.}
	\label{Sensing_service}
\end{figure}

The PMN is distinguished by its ubiquitous sensing capability, making sensing service its core feature while also presenting new design challenges. Essentially, sensing service involves wireless infrastructures, such as BS or roadside units, providing users with state information about the targets of interest through wireless sensing. Fig. \ref{Sensing_service} illustrates various types of sensing service requirements. 

On one hand, users may be interested in their own state parameters, such as position and gesture, which are commonly necessary for localization or fall detection applications. On the other hand, users may request information about other targets from the BS. In such instances, the BS acquires state information about other objects through active sensing and subsequently conveys the data information through the communication function. Thus, the users in PMN can be endowed with the BLoS sensing capabilities. Unlike the communication system, where the primary goal is to transmit information reliably and efficiently, there exist various types of sensing services tailored to specific sensing tasks.      

\begin{itemize}
		
\item \textbf{Detection}: Note that there are diverse types of detection, including symbol detection and fall detection, among others. Strictly speaking, our focus is exclusively on classical radar detection in this context, which involves the identification of the presence of a target within a designated area through echo power detection. This task finds application in various domains, such as unlicensed UAV monitoring and traffic accident warning.

\item \textbf{Estimation}: Localization (or positioning) and tracking represent two fundamental aspects of estimation tasks. Localization involves estimating the state parameters of targets (e.g., range and azimuth), to ascertain their precise spatial coordinates. Tracking expands upon localization by incorporating state prediction and evolution, thereby enhancing sensing capabilities for high-mobility targets. Representative applications encompass autonomous driving and industrial robot control. 

\item \textbf{Recognition}: Recognition involves the extraction of feature information from the targets to identify their category. This feature information may encompass distinctive attributes (e.g., face recognition) or alterations in key parameters (e.g., fall detection). Typically, recognition tasks are accomplished by combining traditional radar signal processing with artificial intelligence algorithms like deep neural network (DNN), providing a higher level of sensing service.         
\end{itemize} 

\subsection{What are the Distinctions and Advantages Over Conventional Cellular Networks?}
Radio resources are valuable assets for operators due to the substantial investment involved in obtaining licenses. From the view of operators, allocating these resources for sensing functions may not be a cost-effective solution. Therefore, it is crucial to clarify the necessity and notable advantages offered by sensing services compared to conventional sensing methodologies.

\begin{itemize}
	
	\item \textbf{Device-Free Sensing}: Current localization technologies in cellular networks (e.g., the global positioning system (GPS)) primarily rely on device-based methods, necessitating signaling devices to be attached to users. This poses challenges in generalizing device-free object sensing scenarios, such as obstacles and pedestrian detection without additional equipment. In contrast, PMN can perform active radar sensing through the ubiquitous deployment of ISAC systems, thereby providing device-free sensing service. 
 
 	\item \textbf{Networked Sensing}: Several state-of-the-art environmental-aware methodologies, including visual systems, vehicular radar, and ultrasonic sensor systems, can also offer device-free sensing and yield good sensing QoS. However, their limited coverage areas result in challenging black zone bottleneck issues. The inherent communication capability of PMN enables networked sensing services through data sharing. With networked sensing, the sensing coverage area can be significantly expanded thanks to spatial diversity. Meanwhile, multi-node collaboration and data fusion techniques can further improve the sensing QoS.   
				 
	\item \textbf{Simultaneous High-Speed Communication and High-Precision Sensing}: As the frequency bands rise into the millimeter wave and terahertz (THz) ranges, it becomes possible to substantially improve communication rates and radio sensing range resolution thanks to the large available bandwidth. Within the context of PMN, both high-speed communication and high-precision sensing can be realized using a single transmission waveform. This is achieved by sharing spectrum resources and signal processing units through appropriate dual-functional signal designs. It is clear that PMN can effectively meet the demands of emerging applications like autonomous driving and human activity recognition.   

\end{itemize}   

 \begin{figure*}[!t]
	\centering
	\includegraphics[width=6.8in]{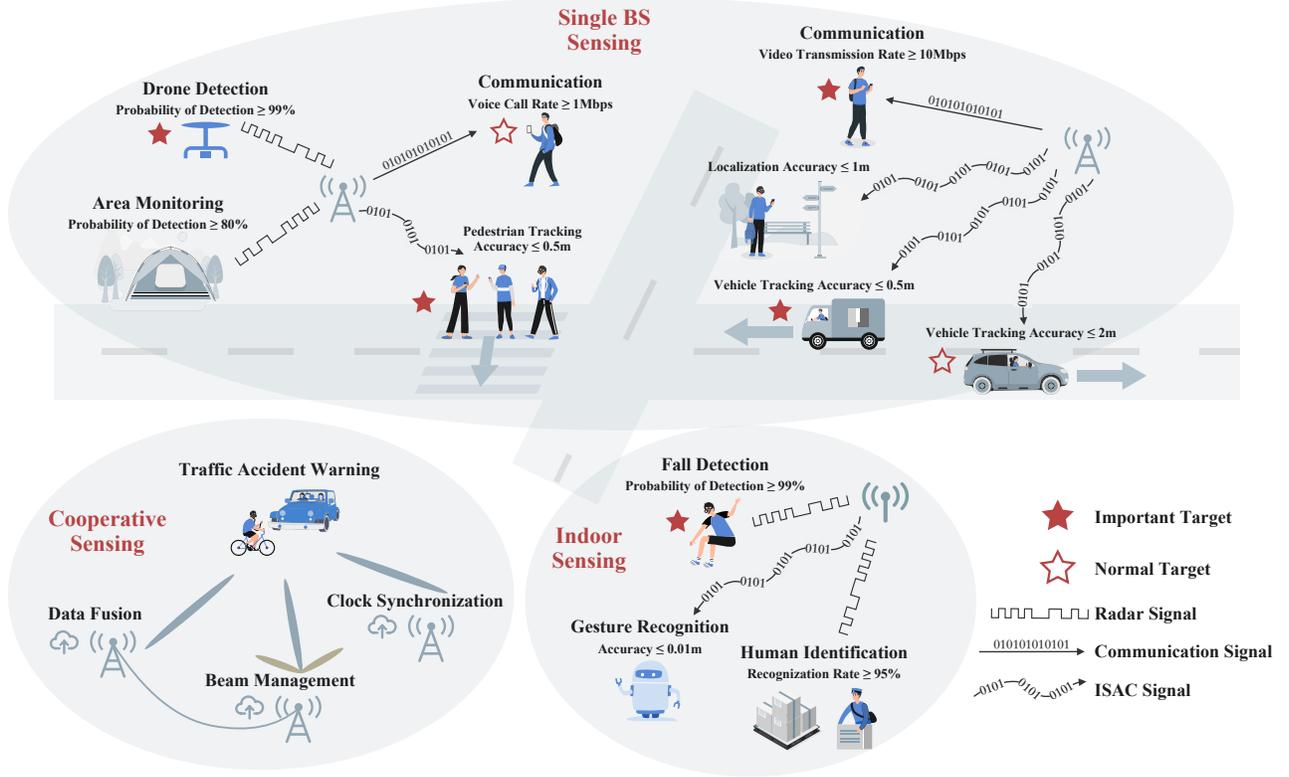}
	\caption{The application scenarios of sensing as a service resource allocation in 6G perceptive mobile networks.}
	\label{system_diagram}
\end{figure*}

\subsection{How to Define Sensing QoS?}

A reasonable and well-defined performance metric for assessing sensing QoS plays a vital role in PMN. This metric serves a dual purpose. Firstly, it serves as a focal point in the optimization of sensing QoS for specific tasks, guiding system design and resource allocation decisions. Secondly, similar to the practice in communication, operators have the potential to implement user-specific sensing QoS levels and levy corresponding fees for these services. Next, we categorize existing metrics into three groups: general metric, MI-based metric, and radar sensing metric.      

\subsubsection{General Metric} This category encompasses metrics that impact all sensing tasks rather than being specific to a particular task. Commonly used metrics in this category include signal-to-interference and noise ratio (SINR), signal-to-clutter and noise ratio (SCNR), and beam pattern matching error (BME). Higher SINR/SCNR indicates a higher proportion of useful signal power, thereby potentially enhancing all types of sensing tasks. Similarly, a desired beam pattern pointing to the target of interest benefits by concentrating energy and reducing clutter interference. The notable advantages of these metrics lie in their applicability to various scenarios and ease of mathematical treatment. However, they may not precisely and clearly characterize the practical sensing performance.      

\subsubsection{MI-based Metric} This category mainly refers to the MI between the received signal and target parameters, measuring the amount of sensing channel information contained in the echo signal. Sensing MI is a ``communication-friendly'' metric with a similar expression to the achievable rate which usually leads to tractable and comparable performance analysis in ISAC systems. Nonetheless, the vague physical definition and lack of operational meanings restrict its practical applications. Recently, within the framework of rate-distortion theory, the authors of \cite{R16} have proved that maximizing MI is completely equivalent to minimizing the MMSE of target impulse response estimation for quadratic Gaussian problems. However, sensing tasks generally focus on the latent parameters (e.g., range and velocity) rather than the sensing channel itself. The connection between the sensing MI of the latent parameters and the corresponding distortions still remains unexplored.

\subsubsection{Radar Sensing Metric}

In contrast to the general metrics and sensing MI, adopting classical radar metrics to measure the performance of diverse sensing tasks may be a more appealing choice. 
\begin{itemize}
	
	\item \textbf{Detection Metric}: Detection involves making binary or multiple decisions to identify whether a target is present or absent. A common metric is the probability of detection (PD), representing the probability of correctly identifying a target. The probability of false alarm (PFA) is another vital metric, which measures the probability of incorrectly declaring the presence of a target when it is in fact absent. In radar applications, the Neyman-Pearson criterion is frequently employed, which maintains PFA below a pre-assigned threshold while maximizing PD.   
	
	\item \textbf{Estimation Metric}: The mean squared error (MSE) between the ground truth and the estimate as a reasonable estimation metric is, however, often challenging to be characterized in practice. Alternatively, the lower bound of the variance for an unbiased estimator, which is referred to as the Cram\'er-Rao bound (CRB), can be employed to measure the estimation QoS. Moreover, in the scenarios of tracking mobile targets, the PCRB becomes a suitable metric by taking into account the Fisher information from both the measured data and prior state models.
	
	\item \textbf{Recognition Metric}: The recognition task can be treated as a classification problem, where the recognition rate evaluates the proportion of the correctly classified instances. Recognition performance not only relies on the classical signal processing approaches, but also heavily depends on the AI algorithm employed. A well-defined recognition metric that is independent of specific operations, is crucial for the integrated sensing, communication, and computing (ISCC) systems.   
	
\end{itemize}

\section{Unified Framework and Unique Challenges}

Fig. \ref{system_diagram} illustrates various application scenarios for sensing-as-a-service within PMN, including single BS-, networked-, and indoor sensing. It is worth noting that with minor modifications, the current cellular network can accommodate such sensing services as an inherent capability in the PMN. For instance, the frame structures and protocols of 5G NR can provide frame-level sensing, while the network architecture of cloud radio access network (C-RAN) has the potential to offer networked sensing \cite{RLiu}. However, there still remain several critical challenges, encompassing the full-duplex hardware configuration, sensing self-interference mitigation, inter-cell interference exploitation, and addressing target returns as outliers in C-RAN, etc. To avoid repetitions, we refer the interested reader to \cite{RLiu} for more details. In this section, we primarily discuss the opportunities and challenges from the perspective of RA schemes that dynamically manage wireless resources in terms of diverse sensing QoS demands.            
     
\subsection{Unified RA Framework}
The served objects can be classified into three categories as follows. (1) \textit{Sensing targets}, which can be either device-free or device-based, represent the entities that need to be sensed. (2) \textit{Communication users} with wireless terminals attached require high-quality communication services exclusively. (3) \textit{ISAC users} requiring both S\&C services, i.e., the intersection of sensing targets and communication users. To fulfill the requirements, BS can adopt either separated S\&C signals or dual-functional signals to optimize the system performance. The RA scheme for ISAC systems involves considering four critical factors as follows.       

\begin{itemize}
	
	\item \textbf{Sensing QoS}: Users in the PMN may have varying sensing QoS requirements for different tasks. Sensing QoS can be evaluated using generalized metrics such as SINR or MI, as well as performance measures for various sensing tasks such as the aforementioned detection, estimation, and recognition metrics.   
	
	\item \textbf{Communication QoS}: In addition to sensing tasks, the PMN may simultaneously provide communication services to users with diverse requirements, such as voice calls and video transmission. Communication QoS is commonly evaluated by the achievable rate and latency for efficiency, as well as the bit error rate (BER) and outage rate for reliability. 
	
	\item \textbf{Resource Budget}: Since resources are constrained in practical applications, resource competition arises among users and S\&C services within the PMN. The allocated resource must not exceed the total available resources in the system design.

	\item \textbf{Priority Level}: Proportional rate constraints are enforced to ensure fairness among users and sensing targets in accordance with different priorities. For instance, important/sensitive targets generally require high sensing QoS to prevent potential accidents, while lower sensing QoS could be acceptable for static and inanimate objects.   
	
\end{itemize} 
      
The PMN allocates a portion of the system resources for sensing targets and the rest for communicating information. In sensing-centric cases, the RA framework aims to optimize the sensing QoS while adhering to the constraints of communication QoS, resource budget, and priority levels. Alternatively, the communication-centric cases may aim to maximize communication QoS or minimize resource consumption, with the other factors acting as constraints based on preset thresholds. 

\subsection{Unique Challenges}

In the conventional individual S\&C system, the purpose of RA is to attain optimal performance trade-offs among the users (or targets) with specific QoS requirements. However, it is worth highlighting that the RA within PMN introduces a fundamental distinction. In this context, in addition to the trade-offs among users, it becomes imperative to consider resource competition and performance \emph{trade-offs between the S\&C services}. This gives rise to unique challenges, as outlined below. 

\subsubsection{Resource Impact Mechanism} 
\begin{figure}[!t]
	\centering
	\includegraphics[width=3.5in]{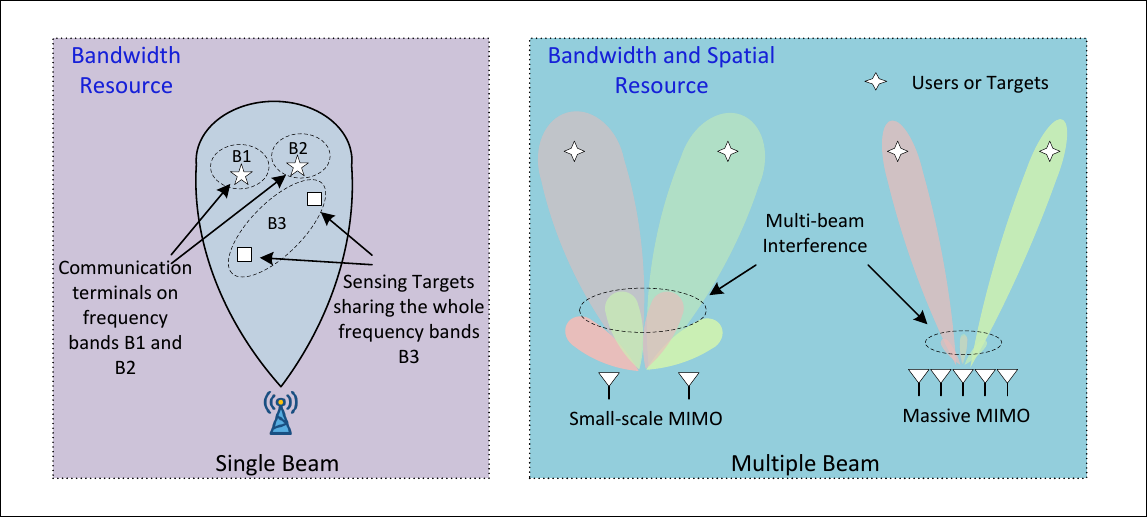}
	\caption{Bandwidth allocation impact mechanism for ISAC system.}
	\label{BAD}
\end{figure}

In Fig. \ref{IR111}, we can observe that the system resources related to S\&C performance can be classified into unique resources (e.g., dwell time for sensing) and shared resources. Note that even though the same resources may have different impacts on S\&C QoSs. Let us take the bandwidth allocation in the beam domain as an example, to illustrate the different resource impact mechanisms between S\&C services, as shown in Fig.\ref{BAD}. 
\begin{itemize}
	\item \textbf{RA in a single beam}: For communication systems, orthogonal bandwidth allocation among the terminals is commonly used to avoid inter-user interference since the terminals are able to operate on heterogeneous frequency bands. By contrast, sensing targets reflect signals at all frequency bands, and thus, all the sensing targets in a single beam have to share the same bandwidth. In other words, bandwidth allocation among users is not required for sensing service in this case. 
	\item \textbf{RA among multiple beams}: The inter-beam interference stems from the power leakage in the overlapped main-lobes or side-lobes. Employing the matched filtering method with orthogonal bandwidth can alleviate inter-beam interference. Additionally, it should be noted that scaling up the number of antennas (spatial resource) sufficiently narrows the formulated beam. In such a scheme, the performance of the overlapped bandwidth allocation scheme may outperform that of the orthogonal allocation scheme thanks to the negligible inter-beam interference, which is however at the price of an enlarged antenna array. 
\end{itemize}   
 
\subsubsection{Resource Impact Scale} The communication metric of achievable rate is inherently a function of the standard bandwidth according to Shannon's theory. In contrast, the sensing metric CRB of the distance estimation does not exhibit a direct correlation with the standard bandwidth. Instead, it is associated with the root-mean-squared (RMS) bandwidth, which is determined by both the standard bandwidth and the frequency-domain waveform \cite{R14}. In other words, the same resource may have different impact scales on S\&C services, which should be primarily considered in RA for ISAC systems. For example, when the BS transmits a perfectly rectangular pulse through a band-limited filter, the RMS bandwidth shows a linear relationship with the standard bandwidth. Nevertheless, it becomes a quadratic relationship when the transmit waveform is a linear frequency modulation wave.   

\subsubsection{Signal Transmission Strategy} The separated S\&C waveform (SW) and the dual-functional waveform (DW) are two representative signaling strategies in ISAC systems, each resulting in different RA approaches. In the SW scheme, S\&C signals are treated as the mutual interference for each service, necessitating the primary focus on mitigation of this interference through RA in such cases. By contrast, mutual interference can be perfectly eliminated by adopting the DW, since the sensing receiver can remove the nuisance data symbols with prior knowledge. This suggests the allocation of overlapping resources for S\&C services within the DW strategy, leading to resource multiplexing gains. Nonetheless, dedicated SW are expected to outperform DW due to their tailored design for respective functionality, such as achieving optimal ambiguous function or maximum channel capacity. Consequently, the selection of the optimal signal strategy becomes a crucial consideration in RA for ISAC systems.

\section{Case Study: Communication-Assisted Sensing}
In PMN, the primary purpose of systems designs shifts from enhancing communication capabilities, as typically seen in conventional networks, to improving the sensing QoS. Users in PMN are expected to attain a significant BLoS sensing capability supported by the data delivery function. As illustrated in Fig. \ref{CAS}, nearby BS with favorable visibility illuminates the targets and captures observations containing relevant parameters through device-free sensing abilities. Subsequently, the BS communicates this sensory information to the users. This approach enables users to plan their routes in advance based on the acquired sensing information. We term this concept as the Communication-Assisted Sensing (CAS) system.

\subsection{Basic CAS Model}

\begin{figure}[!t]
	\centering
	\subfigure[The application scenarios of CAS.]{%
		\includegraphics[width=3.5in]{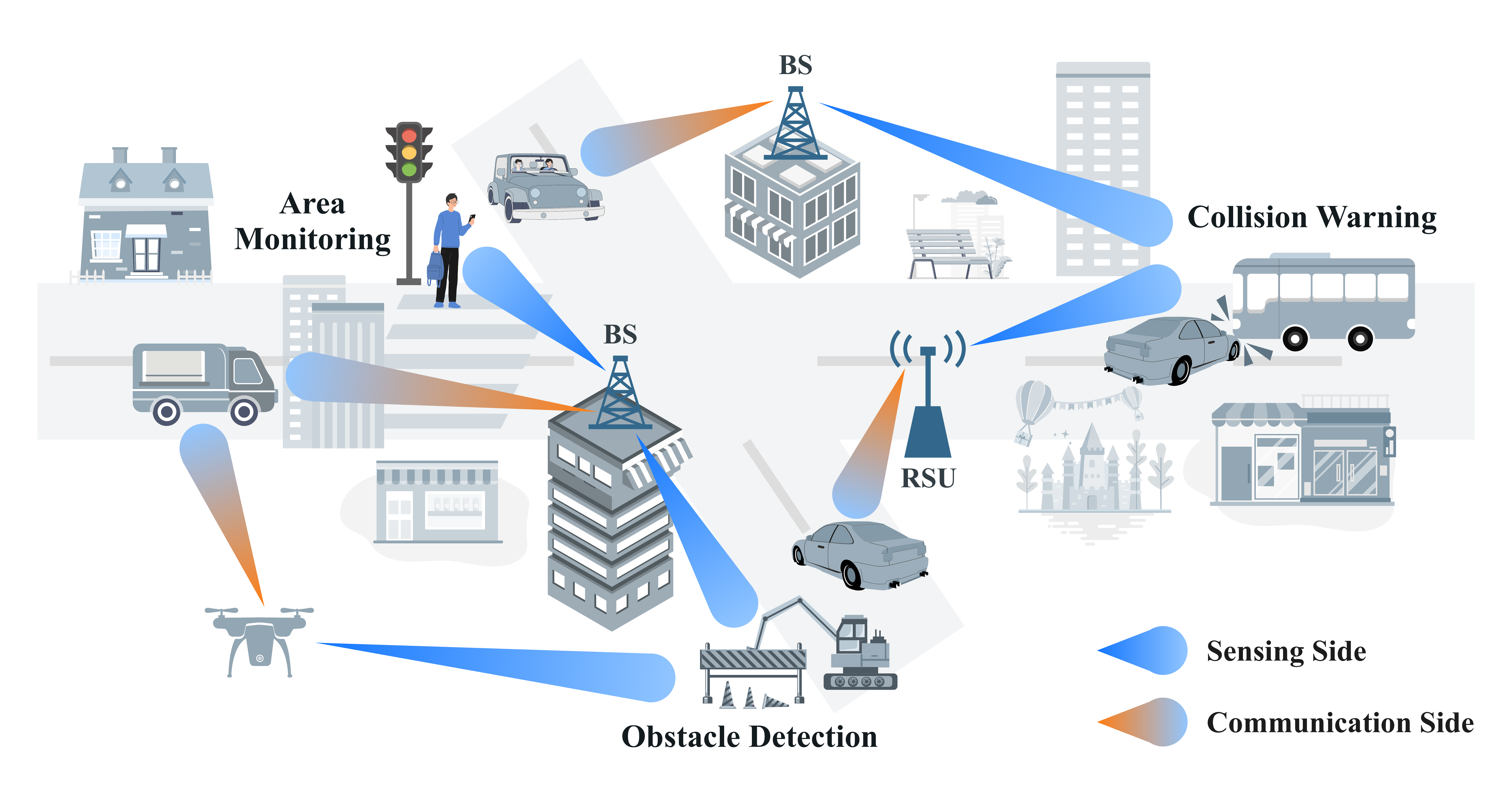}
		\label{CAS}}
	
	\subfigure[The information transmission process.]{%
		\includegraphics[width=3in]{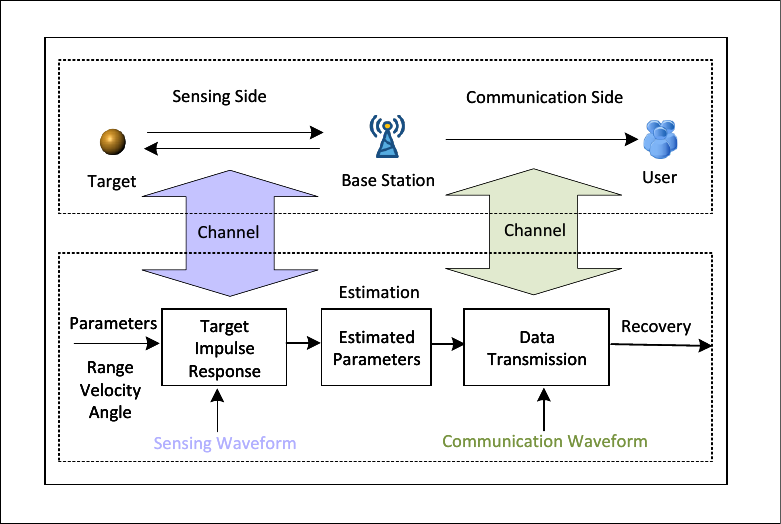}
		\label{CAS_111}}
	
	\subfigure[The power allocation trade-off in separated scheme.]{%
		\includegraphics[width= 2.8in]{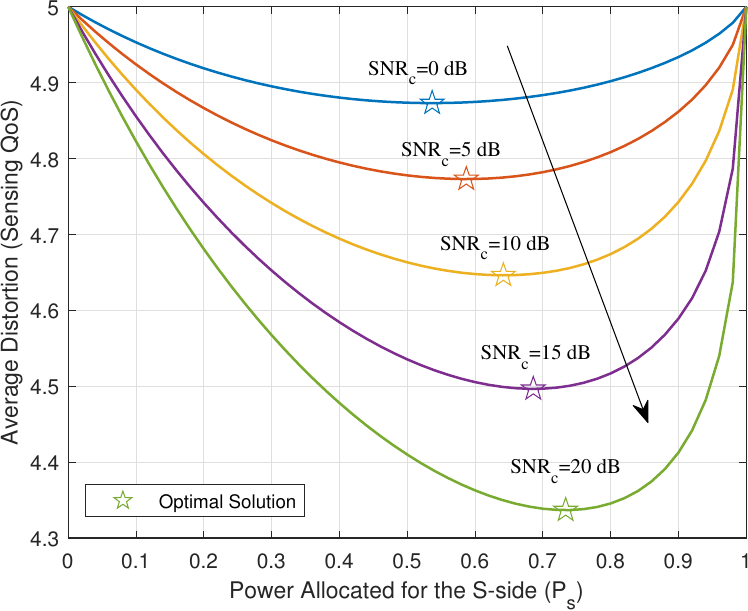}
		\label{CAS_Sim}}
	\caption{The communication-assisted sensing framework.}
\end{figure}

In conventional wireless sensor networks, the communication functions have already been used to deliver the sensory data gathered by other sensors. However, the working pipeline for the CAS framework in PMN differs significantly. Concretely, the S\&C procedures are simultaneously implemented at the BS through sharing the use of the system resources, leading to unique challenges and opportunities in system designs. In summary, the CAS procedure mainly consists of two parts.     

$\bullet$ $\textbf{Sensing side (S-side)}$: The BS transmits sensing waveform to the target and yields the estimated parameter through the noisy received echo signals. 

$\bullet$ $\textbf{Communication side (C-side)}$: The BS transmits the estimated information to the user through source-channel coding, and the user recovers an estimate from the received signal. 

Fig. \ref{CAS_111} demonstrates that the information of the target's parameters successively ``passes through'' the S\&C channels before reaching the user end in the CAS system. The sensing QoS, which can be measured by the distortion between the ground truth and the recovery at the user end, deeply relies on both the S\&C procedures. Consequently, the CAS system exhibits a typical performance trade-off between S\&C procedures.    

\subsection{CAS-based Waveform Design}

Waveform design is a crucial topic within both the domains of radar and communication research. At the S-side, the sensing (estimation) distortion is directly determined by the waveform and the choice of estimator. At the C-side, the communication (recovery) distortion adheres to the source-channel separation theorem (SCT) in lossy data transmission, with the channel capacity depending on the transmitting waveform. Consequently, the CAS-based waveform design is to optimize the sensing QoS (i.e., minimize the sum of S\&C distortions) under the constraints of SCT and resource budget. In what follows, we introduce the waveform designs for the aforementioned two signaling strategies \cite{R17}.     
      
\begin{itemize}
	
	\item \textbf{SW signaling strategy}: In this scheme, the BS transmits individual waveforms for each side, allowing for the utilization of optimal waveform structures. However, this scheme exhibits an obvious resource competition between the S-side and C-side. Fig. \ref{CAS_Sim} illustrates that allocating excessive resources to either side significantly degrades the sensing QoS, thus highlighting a \emph{resource allocation} trade-off.      
		 
	\item \textbf{DW signaling strategy}: In this scheme, the BS transmits a unified waveform that serves both S\&C functions. This strategy capitalizes on resource multiplexing gains and eliminates the resource competition relationship. However, finding a waveform that simultaneously optimizes both S\&C performance is challenging due to the distinct channel states between S-side and C-side, resulting in an \emph{optimal structure trade-off}. 
	
\end{itemize}  

Let us take the waveform design for MIMO systems as an example. In the SW, the eigenspaces of the optimal S\&C waveform can align with their respective channel subspace, while the associated eigenvalues can be determined using the water-filling method. This essentially transforms waveform design into a low-dimensional resource allocation problem for each side. Conversely, in the DW, the eigenspace of a single waveform hardly aligns with both the S\&C channel subspaces simultaneously, which presents a high-dimensional matrix optimization problem. In addition, the CAS framework opens up avenues for a plethora of interesting and challenging research directions, including the CAS for radar parameter estimation with arbitrary prior distributions, networked CAS with task-based quantization, etc.    
       
\section{Open Problems and Future Directions}

The practical network architecture and operational modes for sensing services still require further exploration. Allocating dedicated resources to sensing may sacrifice precious radio resources for communication in mobile networks. Moreover, dual-functional signals carrying confidential data are susceptible to eavesdropping attacks during sensing tasks, potentially giving rise to data security transmission issues. This section elaborates on several open problems and highlights valuable research directions.

\subsection{Fundamental Performance Limits} The resource multiplexing property of the DW schemes enables higher integration gain compared to the SW scheme, especially in the case of low channel quality levels. However, classical theories are insufficient to analyze the performance boundary due to the existence of optimal structure trade-offs. A fundamental problem is to characterize the performance trade-off and coupling mechanism between S\&C tasks while providing a ``theoretic benchmark'' for ISAC system design. To unveil this compromise mechanism, the core idea involves studying the performance limits achievable by the sub-systems, defining unified performance metrics for S\&C, and characterizing their achievable regions \cite{RR15}. This open problem still lacks a clear general conclusion within the academic community.  

\subsection{Non-Orthogonal Resource Allocation}
Although ensuring the orthogonality of resources in time, frequency, and spatial domains can effectively eliminate multi-user interference, it is important to note that orthogonal RA may not always be the optimal choice. Particularly, non-orthogonal RA has the potential to outperform orthogonal RA by compensating for the performance degradation caused by multi-user interference through resource reuse. In addition, leveraging well-studied NOMA techniques may strike a balance between spectral efficiency and user fairness for communications. It is also worth emphasizing that the resources of the DW scheme can be reused without introducing mutual interference. Consequently, the non-orthogonal RA is anticipated to improve system performance through appropriate user grouping/paring.         

\subsection{Networked Sensing Resource Allocation}
In networked sensing, each receiver collects the signals returned from the target and communicates with a fusion center through a backhaul network. This scheme offers several advantages over single BS sensing, primarily due to greater spatial diversity resulting in increased sensitivity and reduced obscuration. However, the networked sensing approach faces three primary challenges. First, clock synchronization is a critical issue in multiple-BS sensing, especially with increasing bandwidths and carrier frequencies. Even a small time offset may lead to significant range error after signal processing. Second, data fusion involves fusing measurements observed by multiple nodes, which is expected to improve the sensing accuracy. Different fusion techniques, such as sending raw data or pre-processed results, directly determine the sensing QoS. Third, when sensing data is collected through wireless communication functions, the data quantization error becomes non-negligible. Similar to the aforementioned CAS techniques, optimizing RA between sensing data acquisition and quantization becomes an intriguing topic.  

\section{Conclusion}\label{Conclusion}
In this article, we systematically examine the concept of sensing-as-a-service and the associated resource allocation (RA) schemes within 6G perceptive mobile networks. We commence with providing an overview of conventional RA frameworks employed in communication, radar, and ISAC systems. Following this, we introduce the concept of sensing-as-a-service and elucidate its distinctions and advantages in comparison to existing cellular networks. Additionally, after defining the sensing quality of service, we elaborate on a unified RA framework, highlighting the unique challenges brought by the trade-off among diverse users and S\&C services. As a further step, we study a representative use case termed as communication-assisted sensing. Finally, we summarize open problems unsolved, highlighting research directions ahead.

%

%

\end{document}